\documentclass{midl} 

\usepackage{tikz}
\usepackage{booktabs}
\usepackage{url}
\usepackage{adjustbox}
\usepackage{mwe} 
\jmlrpages{}
\jmlryear{2019}
\jmlrworkshop{Published as an Extended Abstract in MIDL 2019}
\jmlrproceedings{MIDL}{Medical Imaging with Deep Learning}
\editors{}

\title[E2D Hippocampus Segmentation]{Extended 2D Consensus Hippocampus Segmentation}

 \midlauthor{\Name{Diedre Carmo} \Email{diedre@dca.fee.unicamp.br}\\
   \Name{Leticia Rittner} \Email{lotufo@dca.fee.unicamp.br}\\
   \Name{Roberto Lotufo} \Email{lrittner@dca.fee.unicamp.br}\\
   \addr MICLab, School of Electric and Computer Engineering, University of Campinas, Brazil
   \AND
   \Name{Bruna Silva} \Email{brunaf.10@hotmail.com}\\
   \Name{Clarissa Yasuda} \Email{cyasuda@unicamp.br}\\
   \addr Faculty of Medical Sciences, University of Campinas, Brazil}

\begin{document}

\maketitle

\begin{abstract}
Hippocampus segmentation plays a key role in diagnosing various brain disorders such as Alzheimer's disease, epilepsy, multiple sclerosis, cancer, depression and others. Nowadays, segmentation is still mainly performed manually by specialists. Segmentation done by experts is considered to be a gold-standard when evaluating automated methods, buts it is a time consuming and arduos task, requiring specialized personnel. In recent years, efforts have been made to achieve reliable automated segmentation. For years the best performing authomatic methods were multi atlas based with around 80-85\% Dice coefficient and very time consuming, but machine learning methods are recently rising with promising time and accuracy performance. A method for volumetric hippocampus segmentation is presented, based on the consensus of tri-planar U-Net inspired fully convolutional networks (FCNNs), with some modifications, including residual connections, VGG weight transfers, batch normalization and a patch extraction technique employing data from neighbor patches. A study on the impact of our modifications to the classical U-Net architecture was performed.  Our method achieves cutting edge performance in our dataset, with around 96\% volumetric Dice accuracy in our test data. In a public validation dataset, HARP, we achieve 87.48\% DICE. GPU execution time is in the order of seconds per volume, and source code is publicly available. Also, masks are shown to be similar to other recent state-of-the-art hippocampus segmentation methods in a third dataset, without manual annotations. 

\end{abstract}


\section{Introduction}
Hippocampus segmentation is very important in the diagnosis and treatment of many brain disorders, such as Alzheimer's disease. The hippocampus has an important role in long and short term memory, and many times when affected by some disease gets reduced in size and shape \cite{andersen2007hippocampus}. In epilepsy treatment, in some cases surgical intervention is necessary \cite{yasuda2010dynamic}, and brain MRIs are often used to help in the planning phase. Standard procedure in this cases is to perform a MRI scan of the brain and have experts analyze the shape of the hippocampus. To this day, manual segmentation is still a gold standard, even though interrater variability is a concerning problem \cite{souza2018reliability}. However, manual segmentation still takes time and needs to be performed by specialized personnel.

For some time, the state-of-the-art in automated hippocampus segmentation was composed mainly by methods with execution time in the order of hours per volume computation, with around 0.9 Dice \cite{duarte1999comparison}. Until recently, the most successful methods used the multi atlas approach \cite{wang2013multi}, \cite{iglesias2015multi}, \cite{pipitone2014multi}. In this approach, multiple expert-segmented example images, called atlases, are registered to a target image, and deformed atlas segmentations are combined using label fusion \cite{sabuncu2010generative}. The main drawback of these methods is the time it takes to perform a segmentation, in the order of hours per volume. One notable example is FreeSurfer \cite{fischl2012freesurfer}, a tool with a collection of methods for full brain segmentation that is used nowadays by physicians to aid segmentation, but takes hours to segment a volume. 

Our work is inspired by the need to reduce the computation time of authomatic hippocampus segmentation, using a different approach, namely, Convolutional Neural Networks (CNNs). Recently, some works have also attempted to use CNNs with promising performance \cite{wachinger2018deepnat, thyreau2018segmentation,xie2018near}. Our method consists of evaluating the consensus of volumes generated by three separate Extended 2D (Section \ref{sec:arch}) U-Net like \cite{ronneberger2015u} FCNNs, with encoders initialized in VGG11 \cite{simonyan2014very} and residual connections \cite{he2016deep}. The networks are trained on each brain orientation; sagital, coronal and axial. Our main contribution is a lightweight method with a 200MB memory footprint and 15 seconds mid-range GPU execution time per volume. This method achieves state-of-the-art segmentation performance of 96\% Dice in our test set and employs CNN design ideas and learned knowledge from various works in deep learning, with results visually comparable to other recent hippocampus segmentation methods.

\section{Related Work}

Xie et Gillies (\citeyear{xie2018near}) used a small Convolutional Neural Network (CNN) architecture focused on providing a fast method, which is one of the main advantages of Deep Learning in comparison to previous works. The work focused not only in fast prediction, but also low memory usage and fast training time, which can be difficult to accomplish with CNNs. The author used 2D patches from all three MRI orientations, but fed to a single model that predicts a single voxel classification.

Thyreau et al. (\citeyear{thyreau2018segmentation}), named Hippodeep, is more similar to this work in the sense that it uses Fully Convolutional Neural Networks trained in a region of interest (ROI). However, where we apply one FCNN for each plane of view, Thyreau et al. uses a single FCNN, that starts with a planar analysis swiftly followed by layers of 3D convolutions and shortcut connections. 3D FCNNS are known to be very computational intensive in training due to its large number of parameters, requiring large amounts of data. This study used more than 2000 patients, augmented to around 10000 volumes with augmentation. Initially the model is trained with FreeSurfer segmentations, and later fine tuned using volumes which the author had access to manual segmentations, the gold standard. In our experiments, hippodeep was used for a qualitative analysis of our work. 

Wachinger et al. (\citeyear{wachinger2018deepnat}), named DeepNat, is a whole brain segmentation method that achieves a segmentation of all structures of the brain with around 90\% Dice, including the hippocampus. The method uses 3D patches CNNs to classify voxels and its neighbours, with a multi-task learning strategy. Patches are augmented with coordinates and a novel brain parametrization strategy is presented, to avoid the initial registration problem. Two 3D CNNs are used, first segmenting the background and foreground. Following that, structures on the foreground are segmented with the second 3D CNN.

Although not a hippocampus segmentation work, \cite{lucena2018silver} inspired our consensus strategy that involves the use of three FCNNs performing segmentation over different MRI orientations, merged into a single final volume. While his method uses another network to produce the final consensus, in our post processing we simply add the activation heatmap of each FCNN, apply a pre-defined threshold, and perform 3D labeling to eliminate all but the two bigger connected components. The addition of a fourth network, in our case, did not change results significantly.

\begin{figure}[t]
\floatconts
  {fig:intro}
  {\caption{A coronal slice from a sample of our data. In blue the reference segmentation and in green a correspondent slice of the resulting mask from our method.}}
  {\includegraphics[width=0.8\linewidth]{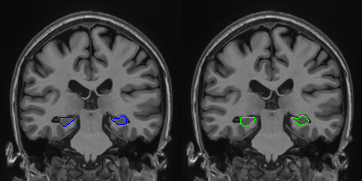}}
\end{figure}
\section{Data and Ethics}
The main, currently private dataset used on this work was collected by medical personnel from the Brazilian Institute of Neuroscience and Neurotechnology (BRAINN) and Hospital HC-UNICAMP. Our dataset contains 214 MNI registered T1 weighted MRI acquisitions made at HC (\figureref{fig:intro}). Almost one third of the acquisitions (66) are from patients that have suffered some modification or removal of one side of to the hippocampus due to surgical treatment of epilpsy. The dataset was originally collected to study volumetry of the hippocampus post surgery.

For this study, hold-out was employed with 80\% for training, 10\% for validation and 10\% for testing. The only pre processing done to our dataset volumes was minmax normalization of int16 values to float16, between 0 and 1. All MRIs have volbrain~\cite{manjon2016volbrain} segmented masks of principal regions of the brain, including the hippocampus. Patients involved on the data acquisition were volunteers and signed consent terms. This research is done in partnership with BRAINN, and this dataset was already used in previous BRAINN research with approval of the UNICAMP Medical Sciences School Ethics and Research Committee (under CEP 1191/2011).

\subsection{CC-359}
As a visual validation and comparison to hippodeep, we used CC-359, a public dataset with 359 volumes, 1.5T and 3T from Siemens and Philips MRI machines \cite{souza2018open}. Contrary to the training data, the volumes are not registered, and have many variations of magnetic field intensity and position of the hippocampus, translated or slightly rotate in relation to MNI registration. This is useful to show if our model is overfitting or not to our MNI registered, more well behaved training data. 

\subsection{HARP}
HARP \cite{frisoni2015eadc} is used as a public validation dataset, constantly used by other works for validation and comparisons. It consists of 135 selected volumes from ADNI, with manual segmentations following the proposed protocol.

\begin{figure}[t]
\floatconts
  {fig:methodology}
  {\caption{An outline of our method. An input volume is analyzed in all three orientations by FCNNs trained in patches over that orientation. Analysis is done in 2D slices and the results are concatenated in a single volume. Following that, our consensus approach and post processing is applied, outputting the final volumetric segmentation.}}
  {\includegraphics[width=0.8\linewidth]{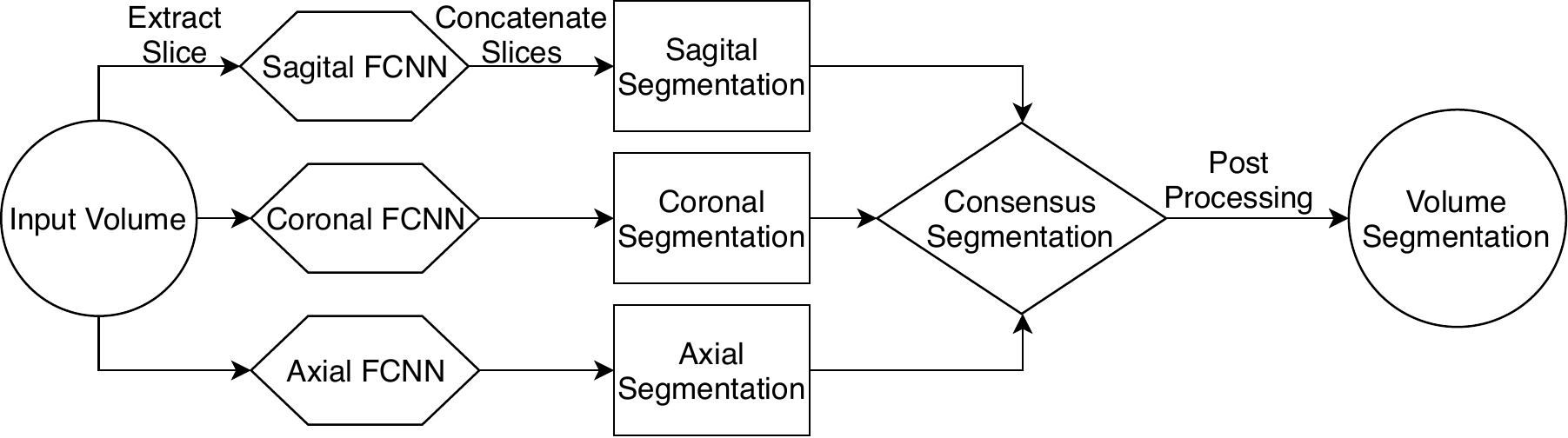}}
\end{figure}

\section{Methodology}

\label{method}
Our analysis consists of three FCNNs examining the brain from three possible orientations, slice per slice, and performing a consensus merging the three volumes generated by each network (\figureref{fig:methodology}). Neighbouring slices are also taken into account on the prediction. The inspiration for this methodology came from the way physicians analyze MRI, using neighbour slices around the point of interest, visualized in all three orientations. Volume segmentation is constructed from running the network multiple times over every slice in the orientation it was trained in. In the following sections, the inner works of our method are described from architecture to final post processing and consensus strategy in more detail. 

\begin{figure}[t]
\floatconts
  {fig:arch}
  {\caption{Diagram showing our architectural choices. Differences from the original U-Net architecure include the 3 channels of grayscale input of neighbour patches, residual connections in the convolution blocks and batch normalization of convolutions in convolution blocks.}}
  {\includegraphics[width=0.8\linewidth]{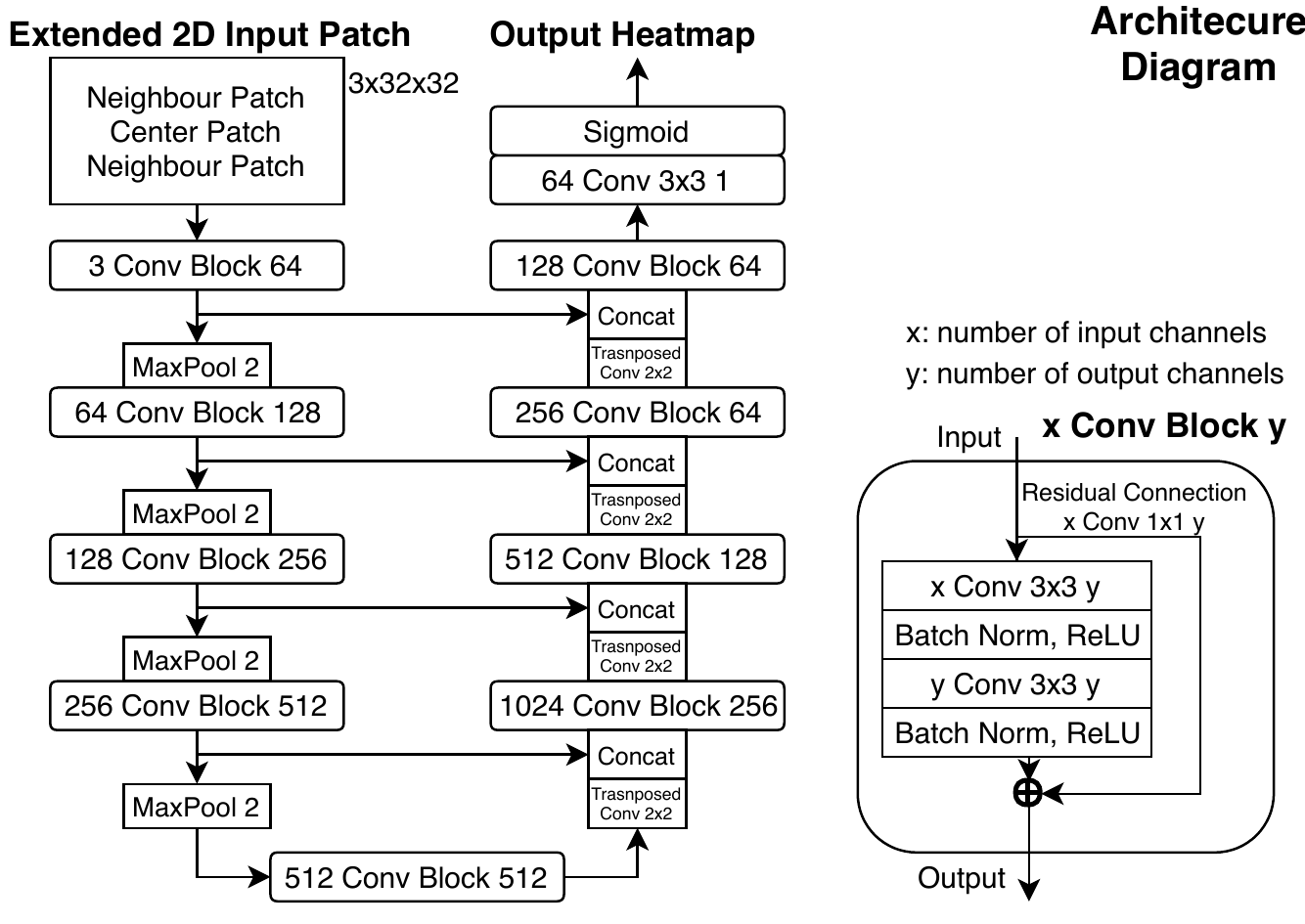}}
\end{figure}

\subsection{Architecture}
\label{sec:arch}
Most of our architectural ideas comes from other successful works. The basic structure of our network (\figureref{fig:arch}) is inspired by U-Net \cite{ronneberger2015u}. However, there are some modifications. Firstly, instead of one single 2D patch as input, two neighbour patches are concatenated leaving the patch corresponding to the target mask in the center. We named that approach as Extended 2D (E2D) for ease of reference. Residual connections based on ResNet \cite{he2016deep} between the input and output of the double convolutional block were added, as 1x1 2D convolutions to account for different number of channels. Batch normalization was added to each convolution inside the convolutional block, to accelerate convergence and facilitate learning \cite{ioffe2015batch}. Also, all convolutions use padding to keep dimensions after the 3x3 convolution and have no bias.  

All previously listed architectural choices improved validation performance on our data and performance on CC-359 (Table \ref{tab:arc}). An attempt was done in using a smaller version of the network, with only 3 max pools and 3 transposed convolutions but convergence was not achieved. Other architectural changes were attempted without success and are not in the scope of this article.



\subsection{Training}
\label{sec:train}
One of the most important steps in achieving good generalization on training CNNs is weight initialization. Poor initialization can have negative impacts in performance. To avoid that, weight transfer from VGG11 is performed, as in \cite{iglovikov2018ternausnet}, to the initial convolutions. Early studies were performed over the validation of the 2D segmentation of each FCNN to determine the best input, loss and learning rate. As an input to the network, an comparison was done between a 128x128 slice center patch and 16x16, 32x32 or a 64x64 random slice patch centered on the hippocampus border. Better validation results were achieved with the 64x64 patch strategy, while the 128x128 center patch resulted in overfitting and the 16x16 patch with a smaller FCNN resulted in under segmentation and more noise. Another early comparison was done between possible loss function choices. Mean Square Error (MSE), Binary Cross Entropy (BCE) and Dice Loss were tested. Better and fast convergence in validation was observed using Dice Loss with a 1.0 smooth factor. Although Dice applies originally to sets, we consider each sigmoid value from 0 to 1 as a set element activation, comparing these values to binary target masks (0 or 1). This allows smooth convergence as the network converges to values close to 1 or 0. Dice Loss is defined as follows:

\begin{equation}
Dice Loss = 1 - \left(\frac{2sum(P*T) + 1}{sum(P) + sum(T) + 1}\right)
\end{equation}

Where P is the prediction sigmoid vector, T is a binary segmentation target vector, sum() denotes the sum of all elements of a vector and * is element wise multiplication. In our experiments, the smooth factor allowed for more stable convergence, with less exploding or vanishing gradients. In training, Dice Loss is calculated per slice and the mean for the mini batch is used. However, when using Dice as a metric for evaluation, Dice is calculated once for the whole final volume without a smooth factor.

With those parameters fixed, a learning rate search for the SGD optimizer with momentum 0.9 was conducted, with the optimal convergence and training speed achieved with 0.001 initial learning rate. The number of epochs was fixed in 500 with 200 patches as mini batch size. After 200 epochs, the learning rate is decayed by a 0.1 factor. While experimenting with parameters and network architecture, the axial orientation was the hardest to learn, often diverging in the middle of training. This makes sense considering the axial orientation is empirically the hardest to identify the hippocampus visually. Adam \cite{kingma2014adam} was attempted as a optimizer but resulted in most times in divergence in the axial orientation. Only states with best validation were saved.

Every patch is generated at runtime. When using the E2D Patch strategy, a patch refers to the center patch and its neighbours. As augmentation, every extracted patch from a slice has a 20\% chance of being of a completely random position on the brain, where the other 80\% are centered on a hippocampus border. There is a 20\% chance of the patch being horizontally fliped. Every patch suffers a variation in brightness between -10 and 10\%, and there is a 20\% chance of gaussian noise with variance 0.0002 and mean 0 being added. It was empirically observed that vertical flips resulted in worst performance. We guess that might be due to the hippocampus being more horizontally than vertically symmetric. Another observation was that noise augmentation helped with generalizing performance to the CC-359 dataset, even though training becomes harder.

\begin{figure}[t]
\centering
\subfigure[]{
	\label{fig:ths}
	\includegraphics[width=0.48\linewidth]{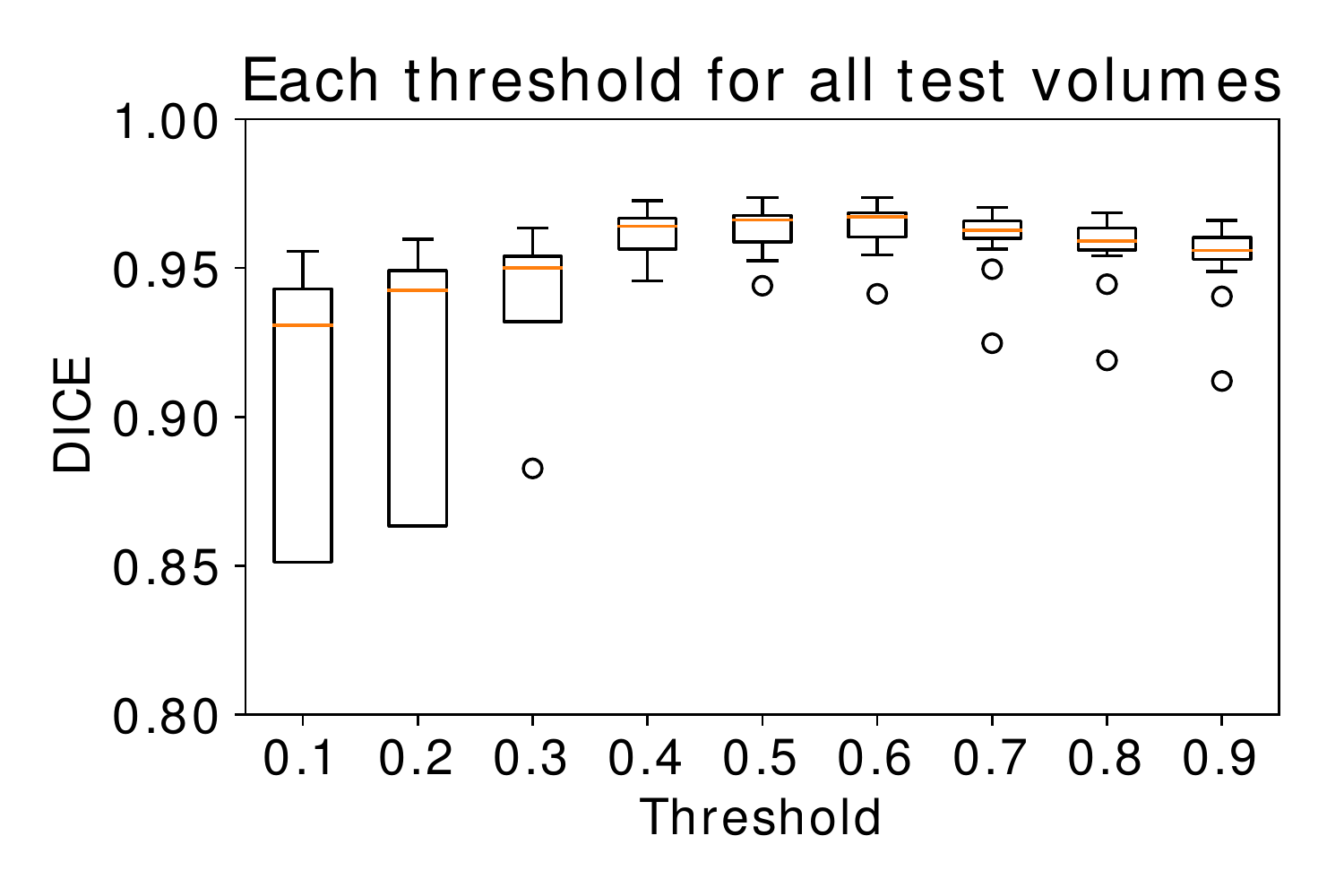}}
\subfigure[]{
	\label{fig:consensus}
	\includegraphics[width=0.48\linewidth]{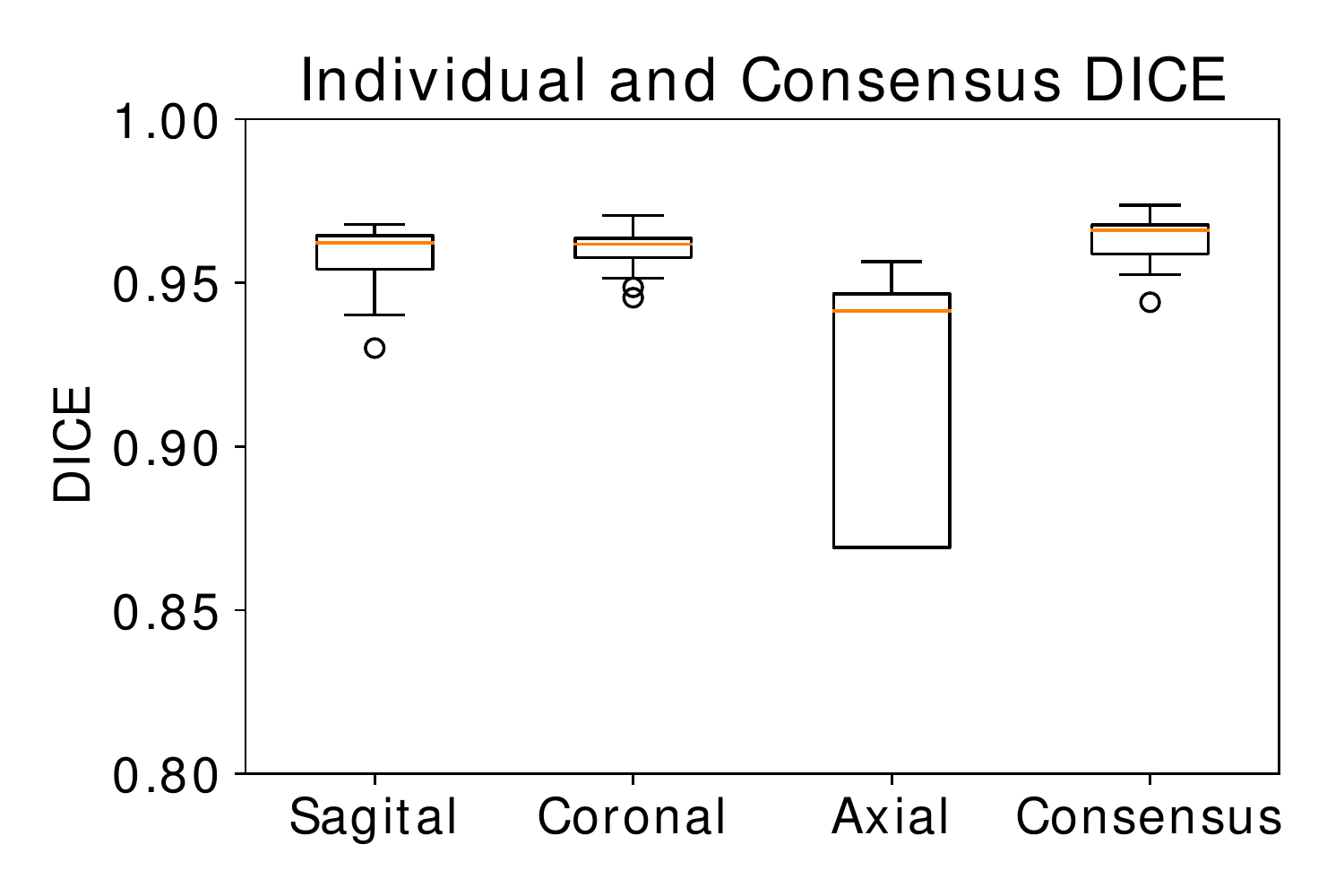}}
\caption{a) Dice values calculated in every binarization threshold (THS) values varying from 0.1 to 0.9 in all 22 test volumes. For all reported Dice results in this paper a THS of 0.5 was used. b) Results considering only one orientation versus the final consensus, using our best model. Consensus displays better performance.}   
\label{fig:post}
\end{figure}

\begin{table}[b!]
\centering
\begin{adjustbox}{max width=\textwidth}
\begin{tabular}{ccccc}
\toprule
\textbf{Test Dice (\%)} & \multicolumn{1}{l}{\textbf{Augmentation}} & \textbf{Residual Connections} & \multicolumn{1}{l}{\textbf{Extended 2D}} & \multicolumn{1}{l}{\textbf{VGG11 Weights}} \\\midrule
92.78                   & -                                         & -                             & -                                        & -                                         \\
93.33                   & \checkmark                                 & -                             & -                                        & -                                        \\
94.81                   & \checkmark                                 & \checkmark                     & -                                        & -                                         \\
95.53                   & \checkmark                                 & \checkmark                     & \checkmark                                & -                                         \\
96.30                   & \checkmark                                 & \checkmark                     & \checkmark                                & \checkmark \\\bottomrule                            
\end{tabular}
\end{adjustbox}
  \caption{Selected final consensus results, showing the improvements on test set volume Dice after including our changes to the U-Net base architecture of each network. Models without E2D have as input a single 32x32 patch in training. Training parameters were fixed as discussed on Section \ref{sec:train}}
  \label{tab:arc}
\end{table}

\subsection{Post-Processing and Evaluation}
After all three networks are trained, in the test phase a volume for each network is generated by segmenting 160x160 center crop slices in their respective orientations. A concatenation of the slices results in one segmentation volume for each network. To generate a final consensus heatmap, each volume is given equal weight of 1/3 and the activation maps are summed. Careful attention was given to the registration process to the final volume to avoid errors, padding the volumes to their original size from the 160x160 center crop. Binarization of the consensus volume is performed with a threshold of 0.5 (\figureref{fig:ths}), and finally, 3D labeling is performed using an implementation from \cite{dougherty2003hands}. The two connected labels with more volume are kept. This post processing raises the performance of our best model by around 10\% in our test set, by removing noise from small false positives in the neck area, skull, and brain ridges and grooves. Also, the consensus has an effect of prioritizing a confident activation from a network (e.g 0.995, or 0.001) instead of uncertain activations, increasing the robustness of the final result (\figureref{fig:consensus}).

\section{Results and Discussion}
\label{sec:res}

\begin{figure}[t]
\floatconts
  {fig:vis}
  {\caption{Visual comparison of masks generated by Hippodeep (blue) and our method (green) in CC-359 data.}}
  {\includegraphics[width=0.9\linewidth]{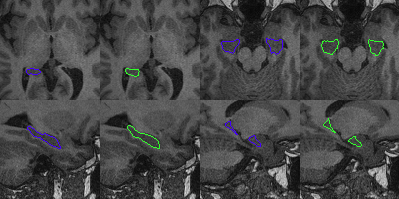}}
\end{figure}



Our results shown state-of-the-art performance in hippocampus segmentation on our test data. Changes listed on Table \ref{tab:arc} on the original U-Net architecture resulted in improvement in our model performance. Also, the consensus strategy resulted in better performance then evaluation following only one orientation,  (\figureref{fig:consensus}). Not using batch normalization resulted in much slower convergence or no convergence at all in most cases.

The first question one asks in front of good results is if the model is overfitting. We report that this method visually generalizes to another, large dataset, CC-359. CC-359 includes different MRI machines and mangetic intensities in relation to our training data. Also, the data in CC-359 is not registered to a common space, and has more neck tissue included. Before the inclusion of our modifications over the U-Nets, the method did not generalized well in CC-359. However, using Dice against Hippodeep \cite{thyreau2018segmentation} masks in CC-359 data, we saw 25\% improvement with residual connections, and 12\% improvement over that with VGG11 weight initialization and the Extended 2D approach. That shows the importance of those modifications on the U-Net architecture for the robustness and generalization of this method. \figureref{fig:vis} shows comparisons of our masks and Hippodeep masks on CC-359 data. Finally, our method used less training volumes than Hippodeep, and runs in around 15 seconds per volume on a mid-range nVidia 1060 GPU.

Additionally, we performed validations in a public dataset, HARP, containing 135 selected subjects from ADNI, comparing to training performed in \cite{isensee2017brain}'s 3D architecture, \cite{thyreau2018segmentation} pre-trained weights predictions and reported results on \cite{ataloglou2019fast}.

\begin{table}[t]
\begin{adjustbox}{center, min width=.8\textwidth}
\begin{tabular}{ccccc}
\toprule
\textbf{Method} & \textbf{Trained on}  & \textbf{\begin{tabular}[c]{@{}c@{}}HARP\\ (DICE \%)\end{tabular}} \\\midrule
\textbf{E2D Consensus} & In-house & 81.39\\
\cite{isensee2017brain} & In-house & 85.86\\
\textbf{E2D Consensus} & In-house + HARP & 87.36\\
\cite{isensee2017brain} & HARP & 86.23\\
\textbf{E2D Consensus} & HARP & \textbf{87.48}\\
\cite{thyreau2018segmentation} & Their data & 85.0\\
\cite{ataloglou2019fast} & HARP & \textbf{90.15}\\
FreeSurver v6.0 & - & 69.8\\
\cite{chincarini2016integrating} & HARP & 85.0\\
\cite{platero2017combining} & HARP & 85.0\\\bottomrule                               
\end{tabular}
\end{adjustbox}
  \caption{This method, named E2D Consensus, shows competitive results on HARP.}
  \label{tab:res}
\end{table}

Table~\ref{tab:res} shows generalization of our method to a public dataset, being competitive with the state-of-the-art. This method achieves 87.48\% DICE when trained on HARP. Interesting to note that, \cite{ataloglou2019fast} is the only method that surpass this method, and it uses a similar initial methodology, using the consensus of CNNs over each volumetric orientation.

Finally, source code is available in an alpha version in \url{https://github.com/dscarmo/e2dhipseg}, with an easy to use interface to run in given volumes. 

\section{Conclusion}
This paper presents a hippocampus segmentation method based on the consensus of three Extended 2D FCNNs with 96.3\% Dice in our test set composed of 22 MRI samples, and 87.48\% DICE in a public validation dataset, HARP. The method also visually displays generalization in another fairly different dataset using hippodeep as a reference. 

Future work will involve acquiring more gold-standard segmentations, and experiments with architecture modifications, to try and improve results on HARP. A follow up paper with additional experiments on Epilepsy is published in~\cite{carmo2020hippocampus}.

\midlacknowledgments{We thank FAPESP for funding this research under grant 2018/00186-0, our partners at BRAINN (FAPESP number 2013/07559-3 and FAPESP 2015/10369-7) for letting us use their dataset on this research and CNPq research funding, process number 311228/2014-3.}

\vspace{1cm}
\bibliography{midl-samplepaper}

\end{document}